# Search for axions in streaming dark matter.

K. Zioutas[1,2], V. Anastassopoulos[2], S. Bertolucci[3], G. Cantatore[4], S.A. Cetin[5], H. Fischer[6], W. Funk[1], A. Gardikiotis[2], D.H.H. Hoffmann[7], S. Hofmann[8], M. Karuza[9], M. Maroudas[2], Y.K. Semertzidis[10], I. Tkatchev[11].

[1] CERN, Geneva, Switzerland, [2] University of Patras, Patras, Greece, [3] INFN, LNF, Bologna, Italy, [4] University and INFN Trieste, Italy [5] Istanbul Bilgi University, Faculty of Engineering and Natural Sciences, Eyup, Istanbul, Turkeyi, [6] University of Freiburg, Germany, [7] TU-Darmstadt, Darmstadt, Germany, [8] Munich, Germany, [9] Department of Physics, Center for micro, nano sciences and technologies, University of Rijeka, Croatia, &, INFN Trieste, Italy, [10] Department of physics, KAIST, &, Center for Axion and Precision Physics Research, IBS, Daejeon, Republic of Korea, [11] INP, Moskau, Russia.

**Abstract:**

A new search strategy for the detection of the elusive dark matter (DM) axion is proposed. The idea is based on streaming DM axions, whose flux might get temporally enormously enhanced due to gravitational lensing. This can happen if the Sun or some planet (including the Moon) is found along the direction of a DM stream propagating towards the Earth location. The experimental requirements to the axion haloscope are a wide-band performance combined with a fast axion rest mass scanning mode, which are feasible. Once both conditions have been implemented in a haloscope, the axion search can continue *parasitically* almost as before. Interestingly, some new DM axion detectors are operating wide-band by default. In order not to miss the actually unpredictable timing of a potential short duration signal, a network of co-ordinated axion antennae is required, preferentially distributed world-wide. The reasoning presented here for the axions applies to some degree also to any other DM candidates like the WIMPs.

Email: zioutas@cern.ch

## 1. Introduction

The theoretically and cosmologically well motivated axion has not been discovered as yet, in spite of the world wide undertaken search lasting already for few decades. In this short note we suggest a new experimental approach for the detection of dark matter (DM) axions, which are expected to occupy the whole of space since the Big Bang. The same reasoning applies also to other DM candidates with similar properties.

The axion haloscope scheme suggested by P. Sikivie [1] is still a widely used method searching for DM axions. This method along with new recent ideas (see presentations in [2]) are searching for this elusive particle by scanning the potential axion rest mass range around $10^{-(4\pm2)}$ eV/c$^2$ by tuning the resonance frequency of the cavity immersed inside the strong magnetic field. For example, the axion haloscope ADMX is presently the only one taking data following this working principle. The width of the resonance is ~1/Q, with Q being the quality factor of the cavity. It is this very narrow resonance response function of the magnetic axion haloscope, which on the one hand optimises its sensitivity and on the other hand increases the scanning time accordingly.

Assuming an externally driven DM flux enhancement occurs, for example, due to temporal gravitational focusing by the solar system bodies, we suggest instead a new detection concept of a broad band axion detection scheme with short scanning period, aiming to improve the axion discovery potential. This work is a summary of a presentation given in 2016 [2].

## 2. The idea

The search for DM, and in particular for relic axions, has been based on the assumed isotropic halo distribution of our Galaxy, with a broad velocity distribution around 240 km/s and an average density of ~0.3 GeV/cm$^3$. This choice might have been the reason behind the failure in detecting the celebrated axion so far.

In this proposal we consider instead possible axion DM streams, which propagate near the ecliptic plane of the solar system. More relevant are of course streams, which get occasionally aligned with the Sun → Earth direction. Because, if such a configuration occurs, the Sun can focus gravitationally low speed ($v$<10$^{-2}$c) incident particles downstream at the position of the Earth [3]. In the ideal case of perfect alignment Stream-Sun-Earth, the axion flux enhancement of the stream can be very large (~10$^6$ [3]) or even much more [4]. It is this temporally axion signal amplification, which (axion) DM searches might utilize. In fact, streaming DM might have a density, which is ~0.3 to 30% of the local mean DM density (~0.3 GeV/cm$^3$). Therefore, even a temporally lensed tiny DM stream propagating along the Sun → Earth direction could still surpass the local DM density. This will give rise to an unexpectedly large DM flux exposure of an axion haloscope like ADMX and the helioscope CAST which has been converted into an axion haloscope (see [5]).

Experimentally, in order to have a real advantage of such burst-like axion flux enhancements due to temporally occurring axion stream alignments, the following two prerequisites are necessary for a relic axion antenna:
a) the covered frequency range must be widened to the maximum, and
b) the scanning time must be shortened as much as possible.

Both features will allow to explore accordingly (gravitationally focused) short axion bursts, though with a decreased axion detection sensitivity. However, if the mentioned large flux (=axion signal) enhancement due to the Sun's gravitational lensing effect can surpass the decreased detection sensitivity, such a scheme will be advantageous for the axion search. This is the proposed alternative detection scheme for the detection of the still elusive axion.

We recall that also planets (including the Moon) work as gravitational lenses for slow speed particles [3]. The flux enhancement at the site of the Earth can be as much as about $10^6$x, which for the case of the Moon can be about $10^4$x for speeds of ~$10^{-4}$c. Due to the planetary motion this opens additional windows of opportunity, i.e., if some axion stream(s) experience a temporal alignment with the direction of a planet-X → Earth. Therefore, it is important to keep large the duty cycle of the fast scanning axion antenna.

## 3. Streaming Dark Matter axions

The co-existence of streaming DM or the galactic dark disk hypothesis have already been discussed (see e.g. [6,7]). Here we only mention a few cases of possible interest. For example, it is widely accepted that the Milky Way disrupts the near Sagittarius (Sgr) Dwarf elliptical Galaxy during its multiple passages through the galactic disk. The tidally stripped stars give rise to the "Sagittarius Stream" with the location of the solar system being probably close to the Sgr-orbits. Therefore, this is expected to result also to streams of DM in our Galaxy and its halo (see Figures 7-10 in [8]). The expected map of our Galaxy by the ongoing *Gaia* mission will probably clarify the predicted Sagittarius (or other) streams. Since we know neither the Sgr original mass nor where it came from [9], simulations are rather indicative.

Another example of streaming axion DM are dense small-scale structures, the axion miniclusters [10]. In fact, in the course of time a fraction of them is disrupted and forms tidal streams, with the axion density being "only" an order of magnitude larger than the average. Stream-crossing events in our neighbourhood might occur only about one in 20 years lasting for a few days. However, a much more favourable scenario arises with axion minicluster(s) being trapped by the solar system during its formation [11]. The expected mass of the axion minicluster is ~$10^{-12}$ $M_\odot$. The estimated maximum axion density enhancement is about $10^5$x the DM average; the Earth crossing time of the very dense axionic clumps is a few days per year. Interestingly, such a trapped cluster mass is below a conservative bound on possible DM density in the Solar System as it is derived from many high precision positional observations of planets and spacecraft [12].

Finally, other (un)predictable streams of DM, including caustics [13], may propagate along a gravitationally favourable direction like one from the Sun or Planet-X towards the Earth, which may further enhance the local flux as it is explained above. An example of potential interest is the alignment (within 5.5°) Galactic Center → Sun → Earth, which repeats annually (18$^{th}$ December). In addition, once in 8-9 years this alignment includes also the "New Moon", i.e., when the Moon is aligned around the same direction towards the Galactic Center and interposed between the Sun and the Earth. In fact, this extended alignment will repeat in 2017, while the last one the same date occurred in 1998.

It is encouraging in this context that since several decades of observations (1935 - ), the Earth's atmosphere shows in December a higher degree of ionisation of otherwise unexplained origin [14]. Interestingly, the ionisation maximum occurs near New Moon, while the active Sun and dynamic ionosphere show also a multifaceted dependence on the longitudinal position of the planets. All these observations lack conventional explanation, and, following ref. [14], they are pointing to overlooked influx of streaming invisible massive matter, which experiences planetary gravitational lensing and/or deflection.

## 6. Conclusion

Compared to the widely assumed isotropic DM distribution, streaming DM axions or other particles with similar properties, may be finally the better source for their discovery. This may result to a large axion signal amplification, which can happen only if streaming DM gets occasionally gravitationally focused at the site of the Earth by the intervening Sun (or a planet) [3]. This is actually equivalent to requiring additionally a planetary fine tuning, making the axion detection even harder.

However, for the reasoning of this work, the claimed streaming invisible matter being at the origin of the observed planetary correlations with the mysterious active Sun and the anomalous high ionization of Earth's atmosphere [14] favour the assumption that axion DM may constitute a component of such (or other) slow speed invisible streams. Then, in such a case, the useful timing of data taking becomes one important factor, though being unknown. We also cannot predict when a DM stream alignment will take place, except, for example, for the 18$^{th}$ December alignment with the Galactic Center.

Therefore, an axion haloscope must cover ideally the whole calendar year. But, taking into account the detection mechanism of an axion haloscope, the other important requirement is a maximum wide-band performance with the fastest possible scanning mode [15]. Since one single axion haloscope cannot fulfill all these requirements, it is obvious that this proposal can be fully realized by a network of haloscopes, preferentially distributed around the Globe, in order to secure a possible discovery. It is interesting for this work that novel axion detection ideas have built-in a very wide-band sensitivity, and such antennae need only to maximize their duty cycle, in order not to miss a signal hidden in a gravitationally squeezed stream.

After all, in spite of the long lasting search for the theoretically and cosmologically well motivated axion, it still remains at large, and therefore, the axion search strategy might need reorientation. This is the purpose of this work.

**Acknowledgments**

We are very thankful to Biju Patla for the support and help we had while preparing this work. His contribution was essential for the completion of this proposal. We also acknowledge the discussion with Alvaro de Rujula; his short critical comment was actually encouraging. One of us (KZ) wishes to thank the organizers of the workshop DARK MATTER AXIONS held in Nordita / Stockholm in December 2016 for the opportunity to present and discuss these ideas, and for the positive response we received.